# Uniaxial to unidirectional transition of perpendicular interlayer coupling in IrMn/CoFe/NiFeO/CoFe quadrilayers

Yu-Jen Wang[1], Chih-Huang Lai[1]*, Po-Hsiang Huang[1], Hsiu-Hau Lin[2], Chih-Ta Shen[1], S. Y. Yang[1], T. S. Chin[1], Tzay-Ming Hong[2], H. J. Lin[3] and C. T. Chen[3]

[1]Department of Materials Science and Engineering, National Tsing Hua University, Hsinchu 30077, Taiwan

[2]Department o f Physics, National Tsing Hua University, Hsinchu 30077, Taiwan

[3]National Synchrotron Radiation Research Center, 101 Hsin-Ann Road, Hsinchu 30077, Taiwan

We studied the interlayer coupling in the quadrilayer consisting of IrMn/CoFe (bottom layer)/NiFeO$_x$/ CoFe (top layer). The in-plane perpendicular interlayer coupling is observed between CoFe layers at room temperature. An anisotropy transition from uniaxial to unidirectional in the perpendicular direction is observed around $T_t$ = 55 K. The nano-oxide layer NiFeO$_x$ shows no distinguishable ferromagnetic signal in the high-temperature (uniaxial) phase, while a strong signal appeared in the low-temperature (unidirectional) phase. A possible two-component scenario, that the nano-oxide layer may contain both amorphous short-range antiferromagnetic domains and superparamagneitc clusters, is proposed to explain the phase transition.

*corresponding author, e-mail address: chlai@mx.nthu.edu.tw

One of the most important developments in magnetic multilayers during the past decades is the interlayer coupling between two ferromagnetic (FM) layers across a nonmagnetic spacer layer. The interlayer coupling through a metallic spacer was found to be oscillatory as a function of the spacer thickness, which was well explained within the RKKY picture.[1,2] For insulating antiferromagnetic (AFM) spacers (NiO[3], and $CoFe_2O_4$[4]), an interesting in-plane perpendicular coupling emerges between adjacent FM layers. Furthermore, an in-plane perpendicular coupling between pinned layers has been observed in the spin valves with the insertion of $NiFeO_x$ or $CoFeO_x$ nano-oxide-layers (NOLs).[5,6] Since these NOLs are very thin and buried in the pinned layers, it is quite hard to characterize/measure their magnetic properties directly and the roles of the NOLs on the interlayer coupling remain unclear. In addition, to our best knowledge, all previous studies on the perpendicular coupling did not explore the global temperature dependence of the interlayer coupling either.

In this Letter, we present the experimental observation of a transition from uniaxial to unidirectional in-plane perpendicular interlayer coupling in the $IrMn/Co_{90}Fe_{10}/NiFeO_x/Co_{90}Fe_{10}$ quadrilayers. We demonstrate. the top CoFe layer develops a uniaxial anisotropy with easy axis perpendicular to that of the bottom CoFe layer at room temperature (RT). As the sample is field-cooled below $T_t \sim 55$ K, the unidirectional anisotropy in the top layer appears. It is rather remarkable that the presence of a thin NiFeO$_x$ NOL, not only mediates the interlayer coupling, it actually goes through a novel phase transition as the thermal fluctuations are reduced and generates a unidirectional anisotropy in the top CoFe layer. We propose a two-component picture to explain our experimental findings.

The multilayer structure Si-substrate/35 Å Ta/12 Å Cu/80 Å Ir$_{20}$Mn$_{80}$/10 Å bottom-Co$_{90}$Fe$_{10}$/21 Å NiFeO$_x$/25 Å top-Co$_{90}$Fe$_{10}$/35 Å Ta was deposited by a sputtering system at a base pressure of 3 X 10$^{-7}$ Torr. The NiFeO$_x$ nano-oxide layers were formed by first depositing Ni$_{80}$Fe$_{20}$ layers and then exposing these layers to oxygen plasma at a partial oxygen pressure of 3 mTorr. The Ta and Cu layers below the IrMn AFM layer were used to promote (111) texture of IrMn, and hence enhanced the exchange-biasing between IrMn and bottom-CoFe. Samples were post-annealed at 200 for 15 minutes in the presence of 1 kOe magnetic field to establish the exchange-biasing of IrMn/bottom-CoFe bilayer. The hysteresis loops, obtained at RT by using a vibrating sample magnetometer (VSM), at 0$^\circ$ (field annealing direction) and 90$^\circ$ are shown in Fig. 1. The presence of NOL completely changes the magnetic property of the top CoFe layer. Fig. 1(a) and inset clearly illustrate that the bottom CoFe is still pinned at 0$^\circ$, but a hard-axis-like sheared loop suggests that the easy axis of the top CoFe is not aligned at 0$^\circ$. A uniaxial anisotropy with easy axis at 90$^\circ$ was observed in the top CoFe layer, shown in Fig. 1(b). These results indicate that the NiFeO$_x$ layer blocks the exchange-biasing from the IrMn layer on the top-CoFe layer, and the magnetization direction of the top CoFe layer seems to be determined by the coupling between NiFeO$_x$ and top-CoFe layers.

To further investigate the origin of in-plane perpendicular coupling, the temperature dependence of the coupling was studied. While the direct hysteresis-loop measurement establishes the magnetization direction of the CoFe layers to be perpendicular at RT, it is rather subtle to observe the temperature dependence of the interlayer coupling by using M-H curves due to complicated layer structures and difficult angular measurements in the setup of the available superconducting quantum interference device (SQUID). On the other hand, magneto-resistance (MR) measurement is very sensitive to relative magnetization orientations of the FM layers adjacent to the Cu spacer. Because the magnetization of the free layer mainly follows the measuring field direction, MR measurement provides robust and precise information about the magnetization orientation of the top CoFe layer. Thus, we grew the full spin-valve structure on top of the quadrilayers to study the temperature dependence of the interlayer coupling. The samples of Si-substrate/35 Å Ta/12 Å Cu/80 Å Ir$_{20}$Mn$_{80}$/ 10 Å bottom-Co$_{90}$Fe$_{10}$/t Å NiFeO$_x$/25 Å top-Co$_{90}$Fe$_{10}$/23 Å Cu/40 Å Co$_{90}$Fe$_{10}$/35 Å Ta (where t is between 15 and 45) were deposited, following the same experimental procedures described in previous paragraphs. MR curves were measured using a physical properties measurement system (PPMS) in the temperature range of 10-300 K.

The spin valves with 21 Å NiFeO$_x$ NOLs exhibit abnormal MR behavior at 0° (direction of the annealing field) as indicated in Fig. 2(a). The exchange bias was eliminated, yielding a low MR ratio of 3.5 % at RT. A pseudo-spin-valve like R-H curve with a large MR value (10.8%) was obtained at 90° as depicted in Fig. 2(b). Since the MR mainly contributes from the top-CoFe/Cu/CoFe (free layer), the observed MR curves imply that the easy axis of the top-CoFe layer is perpendicular to the direction of the annealing field, consistent with the results of hysteresis loops in Fig. 1. For comparison, spin valves without NiFeO$_x$ layers were prepared, and their R-H curve is plotted in the inset of Fig. 2(a). Typical MR curve with ratios of 7.5 % and large exchange biasing fields were obtained at 0°. The increases of the MR ratio in the presence of NOLs can be explained by specular reflection at the interface.[7]

Now we turn to the temperature dependence of the interlayer coupling. During the cooling process, a field of 5000 Oe was applied along 90° (direction perpendicular to the field-annealing for IrMn). The MR curve exhibits a typical pseudo-spin-valve-like loop at temperature range of 60~ 300 K, shown in Fig. 3 (a), revealing uniaxial anisotropy of the top-CoFe layer with the easy axis at 90°. However, asymmetry of the MR curve was evident below 55 K, as shown in Fig. 3(b).

The asymmetry of the MR curve below 55K implies that the anisotropy of top-CoFe layer changes from uniaxial to unidirectional and the top-CoFe layer acquires a finite exchange-bias at 90°. Further reduction of the temperature from 55 K to 10 K increases the exchange-bias field on the top layer from 100 Oe to 630 Oe. A typical exchange-biased MR curve measured at 90° was observed at 10 K, shown in Fig. 3 (c). To make sure the appearance of the finite exchange bias is intrinsic, we reverse the direction of the cooling field and find the direction of the exchange bias in the top CoFe layer changes accordingly. Therefore, through the spin-valve structure, we have demonstrated how the presence of NOL can lead to the biquadratic interlayer coupling at high temperatures and the novel transition from uniaxial to unidirectional anisotropy in the top CoFe layer around $T_t \sim 55$ K.

Where does this biquadratic coupling originate from and why does the suppression of thermal fluctuations lead to the uniaxial-to-unidirectional transition? Although recent experimental investigations have demonstrated 90° interlayer exchange coupling in F/AF/F trilayers (NiFe/NiO/Co[8] or $Fe_3O_4$/NiO/$Fe_3O_4$[3] systems), the NOL in our samples does not seem to have periodic AFM long-range order. Images of high-resolution transmission electron microscope (HRTEM) of the samples containing $NiFeO_x$ show no lattice fringes in the $NiFeO_x$ layer, indicating the $NiFeO_x$ is amorphous without a long-range order. However, to mediate the biquadratic interlayer coupling, only short-range AFM order is necessary. That is to say, if the long-range order in the amorphous NOL is destroyed by the fluctuating short-range AFM domains, the biquadratic coupling can still emerge.

To verify whether the picture of short-range AFM domains makes sense, we estimate the strength of biquadratic coupling and compare with previous experiments. The free energy per unit area of the top-CoFe layer, $E = K_{top}t_{top} \sin^2\Theta - J_2 \sin^2\Theta$, comes from the anisotropy $K_{top}$ associated with field annealing at $0^o$ and the biquadratic coupling $J_2$, where $t_{top}$ is the thickness of the top CoFe layer and $\Theta$ is the angle between the magnetization of the top layer and the annealing field. For the biquadratic coupling to be observed, the condition $J_2 > K_{top}t_{top}$ must be satisfied. In fact, we found that the biquadratic coupling at RT disappeared as the thickness of the top layer exceeded 35 Å. By equating $J_2$ to the anisotropy energy $K_{top}t_{top}*$, where $t_{top}*$ is the maximum thickness (35 Å) that yields the $90^o$ coupling, one obtains the rough estimate of the biquadratic coupling $J_2$. Adapting the commonly used value of $K_{top}$ for Co ($K=5*10^5$ erg/cm$^3$)[9], the estimated $J_2$ is 0.175 erg/cm$^2$, which is of the same order of the magnitude observed in NiFe/NiO/Co systems.[8]

While the short-range AFM domains give nice estimate of the biquadratic coupling with a reasonable magnitude, the transition from uniaxial to unidirectional anisotropies remains unexplained. A better understanding of the NOL is desirable to answer the puzzle. X-ray magnetic circular dichroism (XMCD) was performed on the sample with the film structure of Si-substrate/35Å Ta/12Å Cu/80Å $Ir_{20}Mn_{80}$/10 Å bottom-$Co_{90}Fe_{10}$/21Å $NiFeO_x$/35Å Ta, to reveal the magnetic characteristics of the $NiFeO_x$ layer. The design of the film structure was based on the probing depth of XMCD spectra in total electron yield mode (~80 Å). Figure 4 displays the XMCD spectra in Ni 2p-3d excitation region. Figure 4(a) and 4(b) plot the spectra obtained at RT and at 10 K in a field-cooled (FC) state respectively. The XMCD effect on the Ni L-edges is not obvious at RT, as indicated in Fig. 4(a). The intensity of XMCD effect is proportional to the mean magnetic moment so the weak signal agrees with the picture of amorphous AFM domains in NOL. However, the XMCD signal measured at 10 K, as shown in Fig. 4(b), was considerably enhanced. The significant difference between XMCD measurements at RT and 10 K implies that $NiFeO_x$ layer cannot be explained by short-range AFM domains alone. The data indicate the formation of the residual ferromagnetic moments that become quenched when cooling down. To verify the contributions of "frozen moments", field-cooled (FC) and zero-field-cooled (ZFC) measurements of the magnetization of the (21Å $NiFeO_x$/23Å Cu)$_{10}$ multilayers at different temperatures are performed. The structure of multilayers was chosen to magnify the magnetic signal of $NiFeO_x$. As shown in Fig. 5, the FC and ZFC magnetizations deviate significantly at low temperatures. This provides a direct evidence that the residual magnetic moments exist in NOL and become frozen at low temperature, at which the exchange bias appears along 90° by the cooling field on the top CoFe layer.

Combining the experimental evidences together, we propose a two-component scenario to explain the perpendicular biquadratic interlayer coupling and the transition from uniaxial to unidirectional anisotropy. Above the transition temperature, $T_t$ = 55 K, the NOL mainly consists of short-range AFM domains and superparamagnetic-like components. Since the bottom CoFe layer is firmly pinned by the IrMn layer, it is reasonable to assume the anisotropies of different layers are in the order of $K_{bot}>K_{NOL}>K_{top}$., where $K_{bot}, K_{NOL}$ and $K_{top}$ represent the anisotropy of the bottom CoFe, NOL, and top CoFe layer, respectively. Consider the recent report of Co/NiO/NiFe[8] with the anisotropy order of $K_{Co}>K_{NiO}>K_{NiFe}$, it is observed that the easy axis of the long-range ordered AFM aligns with the easy axis in the bottom Co layer. However, the biquadratic interlayer coupling is found between the intermediate NiO and the top NiFe layers. As we emphasize in previous paragraphs, our experimental data rule out the long-range order in the intermediate NOL due to the amorphous structure. However, the strong pinning from the bottom CoFe layer is likely to align the short-range AFM domains. As a result, we expect the spin structure in NOL is not totally random. Instead, it mainly consists of AFM domains whose spins are aligned along the easy axis of the bottom CoFe layer. Due to the amorphous nature, both types of domains ($0^o$ and $180^o$) exist in NeFeO$_x$ so the short-range AFM domains mediate the perpendicular coupling to the top CoFe layer as in the presence of long-range AFM order.

When cooling down below the frozen temperature, $T_t$, residual magnetic moments start to appear. In addition to the two types of AFM domains, we need the additional superparamagnetic component to explain the experimental observations. This superparamagnetic component may originate from the incomplete oxidation of NiFe nano-clusters. The quenched residual magnetic moments in NOL were aligned by the cooling field at $90^o$ and thus these uncompensated spins define a preferential direction where the unidirectional anisotropy in the top CoFe layer begins to develop.

In conclusion, we observed a transition of 90-degree uniaxial coupling to unidirectional coupling in IrMn/bottom-Co$_{90}$Fe$_{10}$/NiFeOx/top-Co$_{90}$Fe$_{10}$ quadrilayers. NiFeO$_x$ may be composed of short-range AFM domains and superparamagnetic components. At the temperatures above $T_t$, spins of NiFeO$_x$ AFM domains are frustrated along the axis of the exchange-biased bottom-CoFe magnetization. Uniaxial anisotropy perpendicular to the frustrated spin axis was generated on the adjacent top-CoFe layer. When the sample was field cooled to temperatures under 55 K, unidirectional anisotropy arose due to the existence of uncompensated spins originating from superparamagnetic components of the $NiFeO_x$ layer.

This work was supported by the National Science Council of Republic of China under Grant No.NSC-94-2112-M-007-016 and Grant No.NSC-94-2120-M-007-012.

**Figure Caption:**

Fig. 1. Hysteresis loops of the 35 Å Ta /12 Å Cu /80 Å $Ir_{20}Mn_{80}$ /10 Å bottom-$Co_{90}Fe_{10}$ / 21 Å $NiFeO_x$ /25 Å top-$Co_{90}Fe_{10}$ /35 Å Ta multilayer with the field applied along (a) 0°, and (b) 90°. The inset figures show the hysteresis loops at a low field region.

Fig. 2. Room temperature R-H curves of the spin valves 35 Å Ta/12 Å Cu/80 Å $Ir_{20}Mn_{80}$/10 Å bottom-$Co_{90}Fe_{10}$/21 Å $NiFeO_x$/25 Å top-$Co_{90}Fe_{10}$/23 Å Cu /40 Å $Co_{90}Fe_{10}$ /35 Å Ta with the applied field along (a) 0°, and (b) 90°. The inset is the R-H loop of the spin valve without $NiFeO_x$ layer.

Fig. 3. MR curves of the spin valves 35 Å Ta/12 Å Cu/80 Å $Ir_{20}Mn_{80}$/10 Å bottom-$Co_{90}Fe_{10}$/21 Å $NiFeO_x$/25 Å top-$Co_{90}Fe_{10}$/23 Å Cu/40 Å $Co_{90}Fe_{10}$ /35 Å Ta with the applied field at 90° measured at (a) 60 K (b) 55 K and (c) 10 K.

Fig. 4. XMCD spectra at the Ni $L_{2,3}$ edges for the $NiFeO_x$ layer of the multilayer 35 Å Ta /12 Å Cu /80 Å $Ir_{20}Mn_{80}$ /10 Å bottom-$Co_{90}Fe_{10}$/21 Å $NiFeO_x$ /35 Å Ta measured at (a) RT and (b) 10 K. The XMCD spectra were taken with the projection of the spin of incident photons parallel ($\rho^+$, solid curve) and antiparalle ($\rho^-$, dashed curve) to the spin of the Ni 3d majority electrons. $\rho^+$ -$\rho^-$ represents the signal of XMCD effect.

Fig. 5. Temperature dependence of magnetization of (21Å NiFeO$_x$/23Å Cu)$_{10}$ multilayers by using zero-field-cooling (ZFC) and field-cooling (FC).

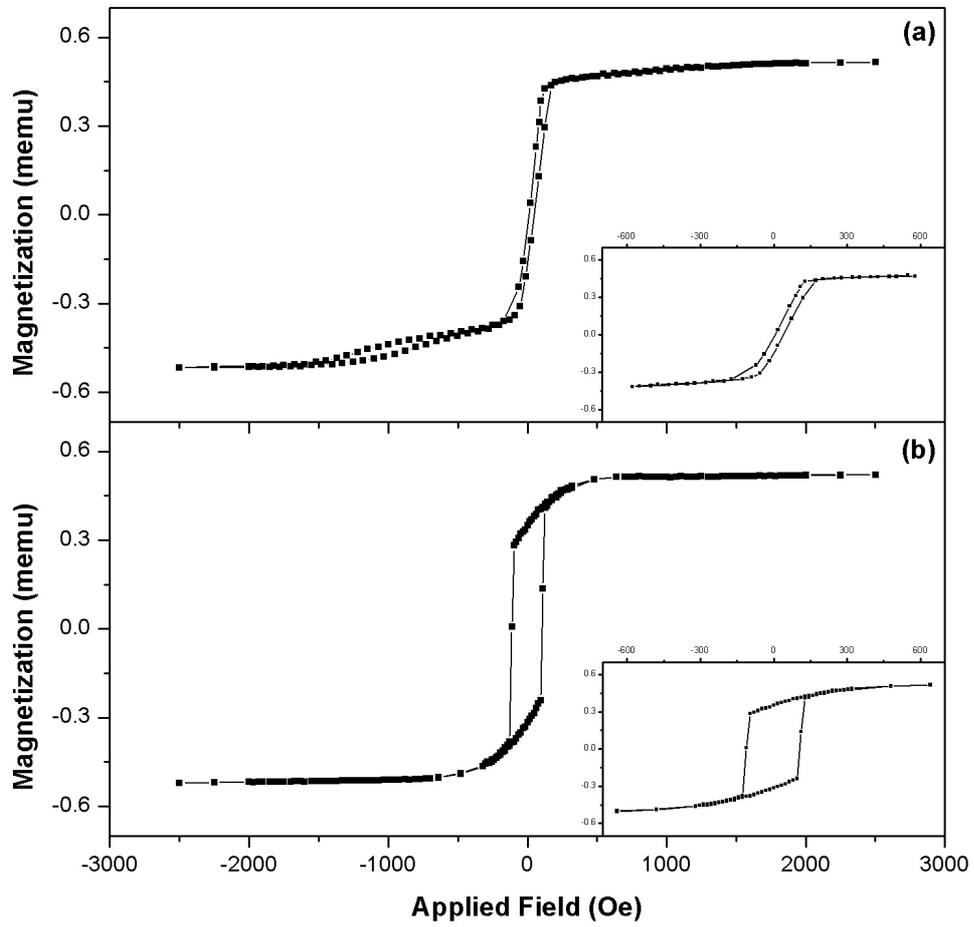

Figure 1, Yu-Jen Wang, APL

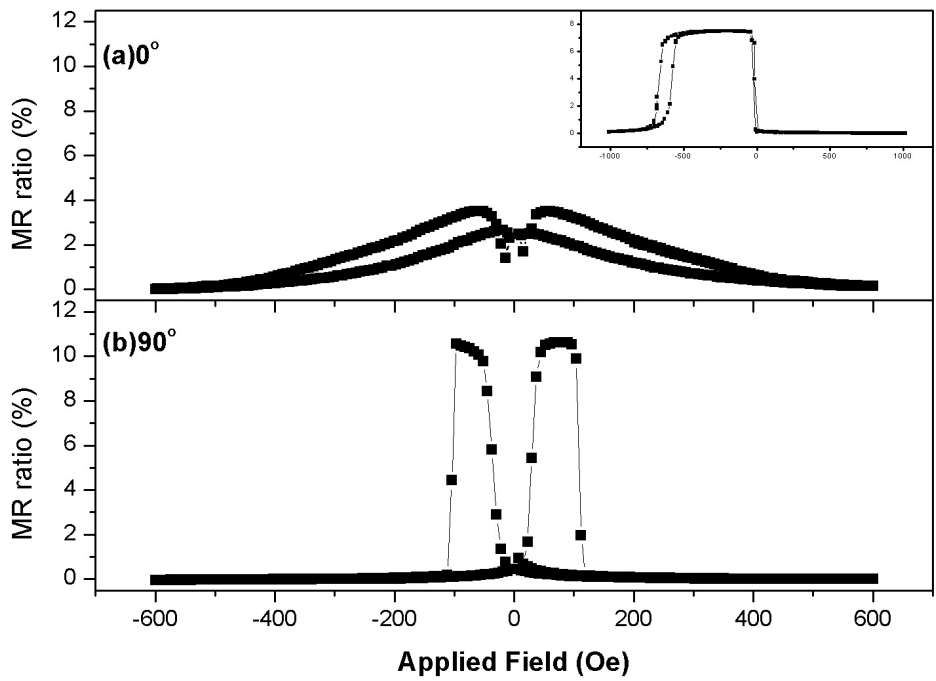

Figure 2, Yu-Jen Wang, APL

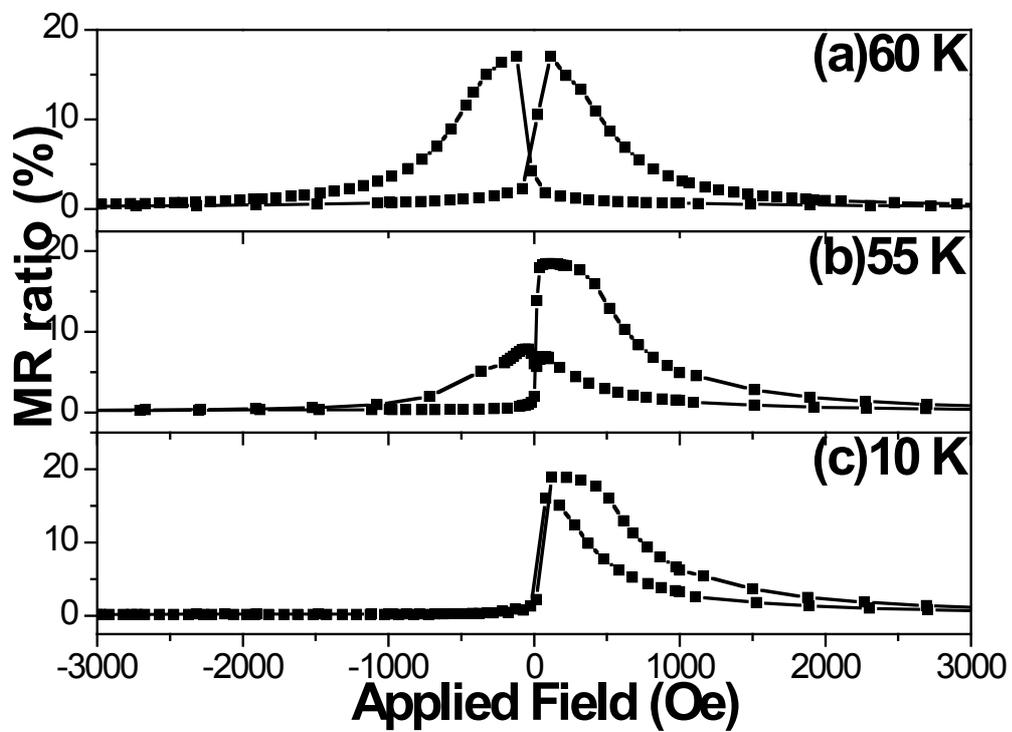

Figure 3, Yu-Jen Wang,
APL

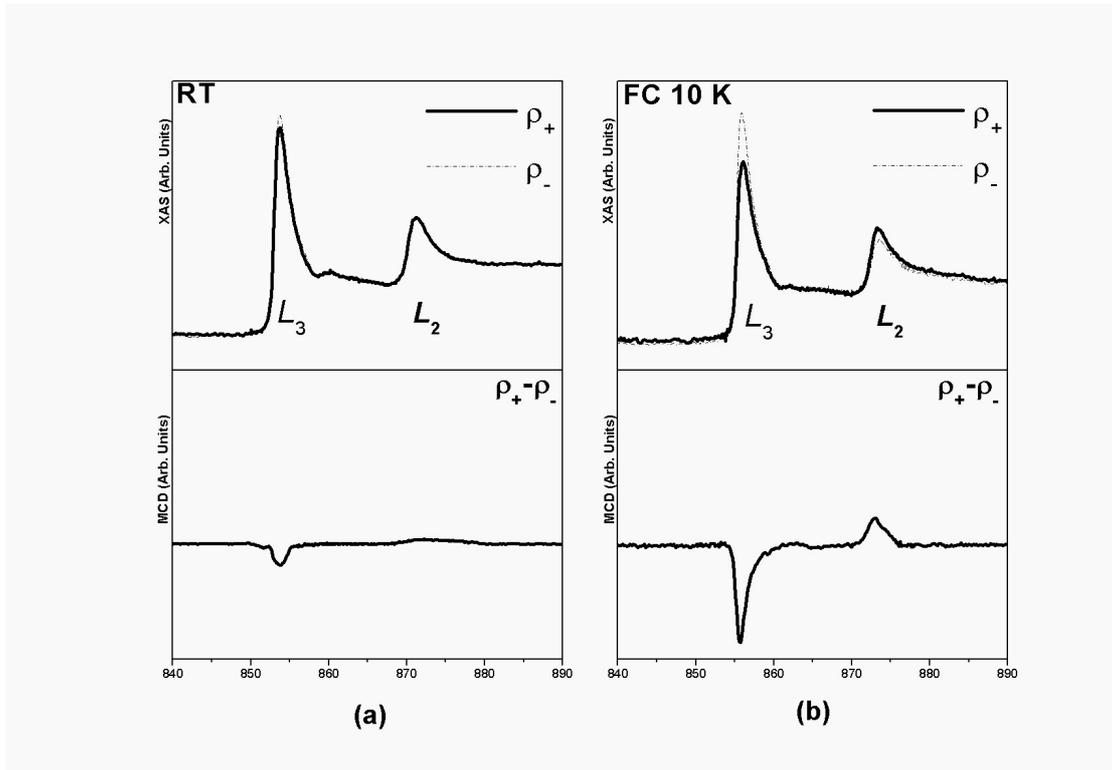

Figure 4, Yu-Jen Wang, APL

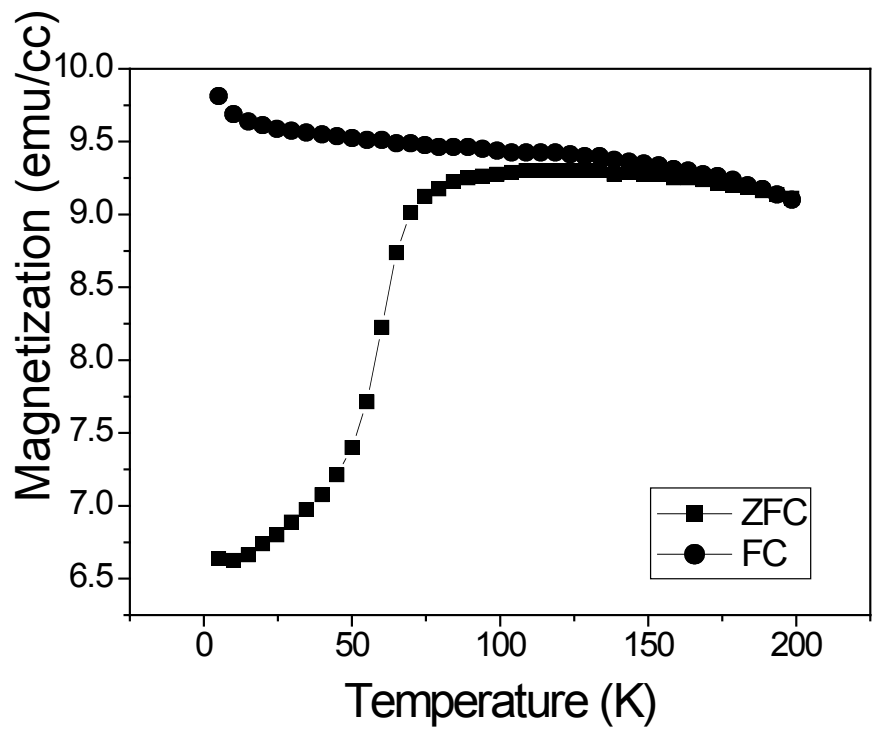

Figure 5, Yu-Jen Wang, APL